\documentclass[aps,prl,twocolumn,showpacs,amsmath,amssymb]{revtex4-1}
\usepackage[pdfusetitle,pdfauthor={Navdeep Rana}]{hyperref}
\usepackage[dvipsnames]{xcolor}
\usepackage[final]{graphicx}
\usepackage{bm}
\usepackage{enumerate}
\usepackage[capitalize]{cleveref}
\crefname{equation}{Eq.}{}
\crefrangeformat{equation}{Eqs.~(#3#1#4)-(#5#2#6)}
\usepackage{tikz}

\newcommand{\Tr}{\mathrm{Tr}}

\newcommand{\Rey}{\mbox{Re}}
\newcommand{\Cn}{\mbox{Cn}}
\newcommand{\vel}{\bm{u}}

\newcommand{\highre}{\Rey=\pi\times10^{4}}
\newcommand{\lowre}{\Rey=0}

\newcommand{\REM}[1]{{}}

\begin{document}
\title{Phase ordering, topological defects, and turbulence in the 3D incompressible Toner-Tu equation}
\author{Navdeep Rana}
\author{Prasad Perlekar}
\affiliation{Tata Institute of Fundamental Research, Centre for Interdisciplinary Sciences, Hyderabad, India}

\begin{abstract}
    We investigate phase ordering dynamics of the incompressible Toner-Tu equation in three dimensions. We show that the
    phase ordering proceeds via defect merger events and the  dynamics is controlled by the Reynolds number $\Rey$.  At
    low $\Rey$, the dynamics is  similar to that of the Ginzburg-Landau equation.  At high $\Rey$, turbulence controls
    phase ordering. In particular, we observe a forward energy cascade from the coarsening length scale to the
    dissipation scale.
\end{abstract}

\maketitle

Phase ordering (or coarsening) refers to the dynamics of a system from a disordered state to an orientationally ordered
phase with broken symmetry on a sudden change of the control parameter~\cite{bray1993,chaikin1995}. Biological systems
such as a fish school or a bird flock show collective behavior - an otherwise randomly moving group of organisms start
to perform coherent motion to generate spectacularly ordered patterns whose size is much larger than an individual
organism \cite{ramaswamy2010,marchetti2013,ramaswamy2019}. Although the exact biological or environmental factors that
trigger such transition depend on the particular species, physicists have successfully used the theory of dry-active
matter to study disorder-order phase transition in these systems \cite{toner1995,chate2020}. Theoretical studies have
revealed that for an incompressible flock, where we can ignore density fluctuations, the order-disorder phase transition
is continuous \cite{chen2015,chen2016,chen2018}. 

In classical spin systems with continuous symmetry, domain walls or topological defects are crucial to the growth of
order and the existence of equilibrium phase transition~\cite{bray1993,chaikin1995}. Several studies have highlighted
the role of defects in the phase-ordering dynamics in spin systems \cite{ostlund1981,toyoki1990,lau1989,sinha2010}.
Interestingly, suppression of defects in the two-dimensional (2D) XY model~\cite{sinha2010} and the three-dimensional
(3D) Heisenberg model~\cite{lau1989} destroys the underlying phase-transition, and the system remains ordered at all
temperatures.

Phase-ordering has been investigated in dry, polar active matter~\cite{mishra2010,das2018,rana2020,chardac2021}.
However, only recent studies have started to explore the role of defects in phase ordering.  We investigated
phase-ordering in the two-dimensional (2D) incompressible Toner-Tu (ITT) equation in an earlier study \cite{rana2020}.
Our study revealed that the phase-ordering proceeds via coarsening of defect structures (vortices), similar to the
planar XY model. However, the merger dynamics had similarities with vortex mergers in 2D fluid flows. More recently,
experiments \cite{chardac2021} investigated coarsening dynamics in 2D dry-active matter and observed, consistent with
Ref.~\cite{rana2020}, that phase-ordering proceeds via the merger of topological defects. 

How does phase ordering proceeds in three-dimensional (3D) polar, dry-active matter? In this paper, we investigate this
question by performing high-resolution direct numerical simulations of the incompressible Toner-Tu (ITT) equations. We
show that phase ordering in the 3D ITT equation proceeds via defect merger. The Reynolds number $\Rey$ - the ratio of
inertial to viscous forces - controls the merger dynamics. For small $\Rey \to 0$, the ordering dynamics have
similarities with the three-dimensional Heisenberg model. For large $\Rey$, on the other hand,  turbulence drives the
evolution and speeds up phase ordering. In particular, we observe an inertial range with Kolmogorov scaling in the
energy spectrum and a positive energy flux \cite{frisch1995}. 

The 3D incompressible Toner-Tu (ITT) equation is \cite{chen2016}
\begin{equation}\label{eq:tonertu}\begin{aligned}
    \partial_{t} \bm{u}+\lambda\bm{u}\cdot\nabla\bm{u}&= -\nabla P + \nu\nabla^2 \bm{u} + \bm{f}.
\end{aligned} \end{equation}
Here $\vel({\bm x},t)\equiv(u_x,u_y,u_z)$, and $P({\bm x},t)$ are the velocity and the pressure fields, $\bm{f} \equiv
\left(\alpha - \beta |\bm{u}|^2\right)\bm{u}$ is the active driving term with coefficients $\alpha, \beta>0$, $\lambda$
is the advection coefficient, and $\nu$ is the viscosity. The incompressibility constraint $\nabla \cdot {\bm u}=0$
relates the velocity to the pressure. Note that $u=0$ and $u=U$ with $U\equiv\sqrt{\alpha/\beta}$ are the unstable and
stable homogeneous solutions of the ITT equation. As we are interested in phase-ordering under a sudden quench from a
disordered configuration to zero noise, we do not consider a random driving term in \cref{eq:tonertu}. Note that for
$\beta=0$ and $\lambda=1$, \cref{eq:tonertu} reduces to the linearly forced Navier-Stokes equation \cite{rosales2005},
whereas it reduces to the Ginzburg-Landau (GL) equation~\cite{onuki2002} for $\lambda=0$ and in the absence of pressure
term. Therefore, similar to the GL equation we expect topological defects to play crucial role in phase-ordering
dynamics of the ITT equation. 

By rescaling the space $x\to x/L$, the time $t\to\alpha t$, the pressure $P \to P/\alpha L U$, and the velocity field $u
\to u/U$ in \cref{eq:tonertu}, we obtain the following dimensionless form of the ITT equation \cite{rana2020}
\begin{equation} \label{eq:nditt} \begin{aligned}
    \partial_{t}\vel + \Rey \Cn^{2}\vel\cdot\nabla\vel = -\nabla P + \Cn^{2}\nabla^2 \vel + \left(1-|\vel|^{2}\right)\vel,
\end{aligned} \end{equation}
where $\Rey\equiv \lambda U L/\nu$ is the Reynolds number with $U\equiv\sqrt{\alpha/\beta}$, and $\Cn=\sqrt{\nu/\alpha
L^2}$ is the Cahn number.

We use a pseudospectral method to perform direct numerical simulation (DNS) of \cref{eq:nditt} in a tri-periodic cubic
box of length $L=2\pi$. Wediscretize the box with $N^3$ collocation points with $N=1024$ and use a second-order
exponential scheme for time integration \cite{cox2002}. We decompose the velocity field into its mean ${\bm
V}(t)=\langle \vel \rangle$ and fluctuating part ${\bm u}'(x,t)\equiv \vel(x,t) - \bm{V}(t)$ to investigate the ordering
dynamics, where the angular brackets denote spatial averaging.  Along with the velocity field, we monitor the evolution
of the energy spectrum
\begin{equation}\label{eq:spectrum}\begin{aligned}
    E_k(t) \equiv \frac{1}{2}\sum_{k-\frac{1}{2}\leq p < k+\frac{1}{2}}|\widehat{\vel^\prime}_{\bm{p}}(t)|^2,
\end{aligned}\end{equation}
and the total energy is $E(t) = \frac{1}{2}V^2(t) +{\mathcal E}(t)$.  Here, $\widehat{\vel}_{\bm{k}}(t)\equiv
\sum_{\bm{x}} \vel(\bm{x},t) \exp(-i \bm{k}\cdot\bm{x})$ is the Fourier transform of the velocity field with
$i=\sqrt{-1}$~\cite{rana2020}.  The flow is initialized with a disordered configuration with $E_k(t=0)=A k^2$ and
$A=10^{-8}$. In our DNS, we choose $\Cn=10^{-2}/2\pi$, $\alpha=1$, $\beta=1$, and investigate phase ordering for
different values of $\Rey$.

\begin{figure}[!ht]
    \centering
    \includegraphics[width=\linewidth]{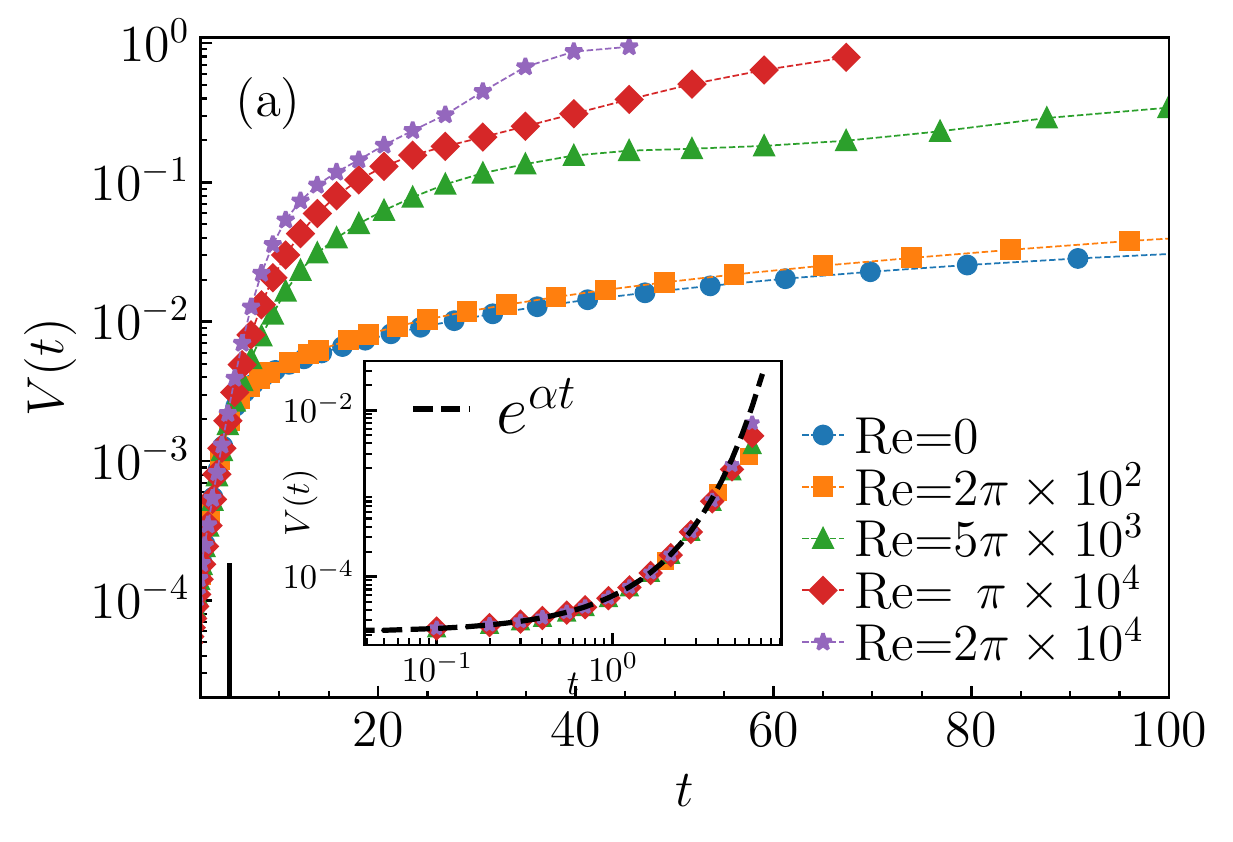}
    \includegraphics[width=\linewidth]{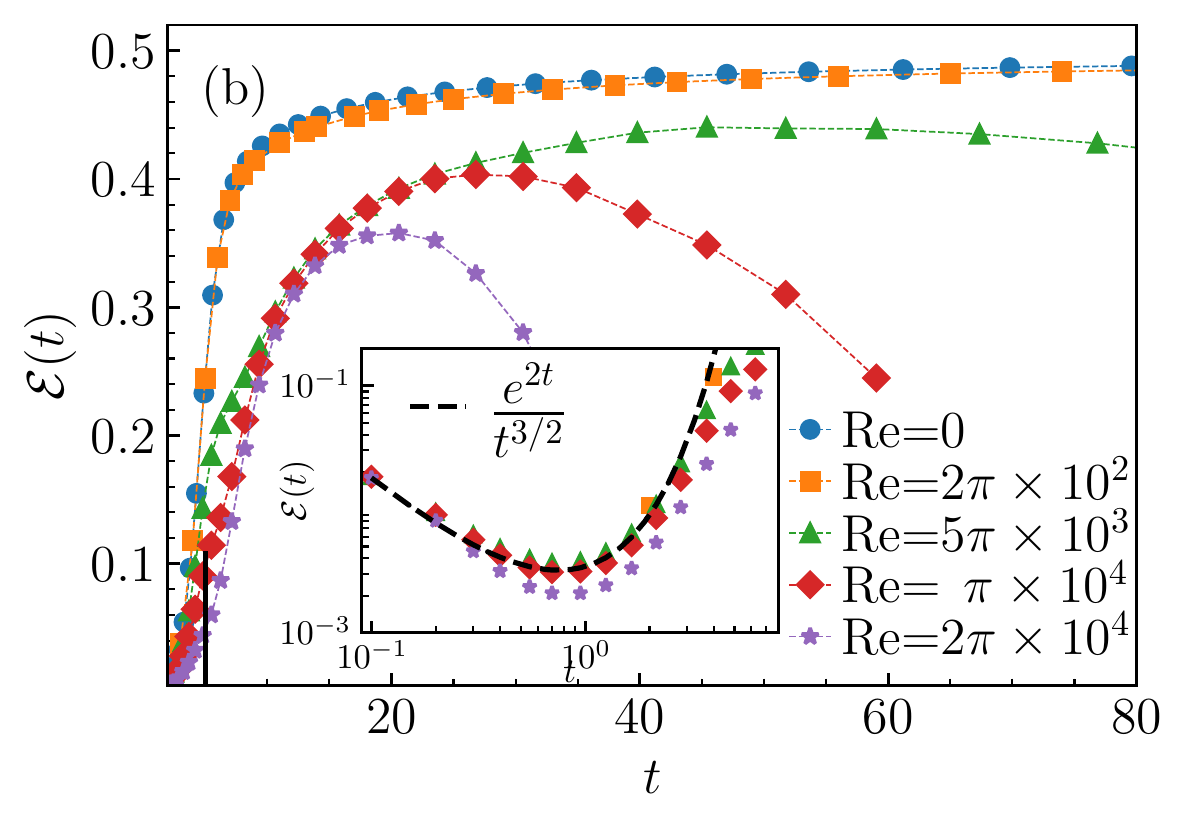}
    \caption{\label{fig:order}
        Time evolution of (a) $V(t)$ and (b) ${\mathcal E}(t)$ for different $\Rey$.  (Insets) Zoomed in plots showing
        early time-evolution along with the theoretical prediction (dashed line).
    }
\end{figure}

In \cref{fig:order}, we plot the magnitude of the mean velocity $V(t)$ and the fluctuation energy ${\mathcal
E}(t)=\sum_{k=1}^{N/2} E_k(t)$.

At early times, the nonlinearities in \cref{eq:tonertu} can be ignored and we arrive at the following time evolution
equation of the energy spectrum \cite{bratanov2015,rana2020}
\begin{equation}\label{eq:gaussian}\begin{aligned}
    \partial_t E_k(t) \approx 2(\alpha - \nu k^2) E_k(t).
\end{aligned}\end{equation}
Using \cref{eq:gaussian} and the initial condition for the energy spectrum we obtain $V(t) \sim \exp(\alpha t)$ and
${\mathcal E} \sim \exp(2 t)/t^{3/2}$ [see \cref{fig:order}].

The departure from the early exponential growth of $V(t)$ and ${\mathcal E}(t)$ marks the onset of the phase-ordering
regime. We observe that $V(t)$ approaches the ordered state faster by increasing the Reynolds number. On the other hand,
${\mathcal E}(t)$ first increases, attains a plateau and then decreases. The plateau region and the peak value of
${\mathcal E}(t)$ decrease with increasing Reynolds. Later we show that the width of the plateau region is related to
strength of the energy cascade.

\begin{figure*}
    \centering
    \includegraphics[width=\linewidth]{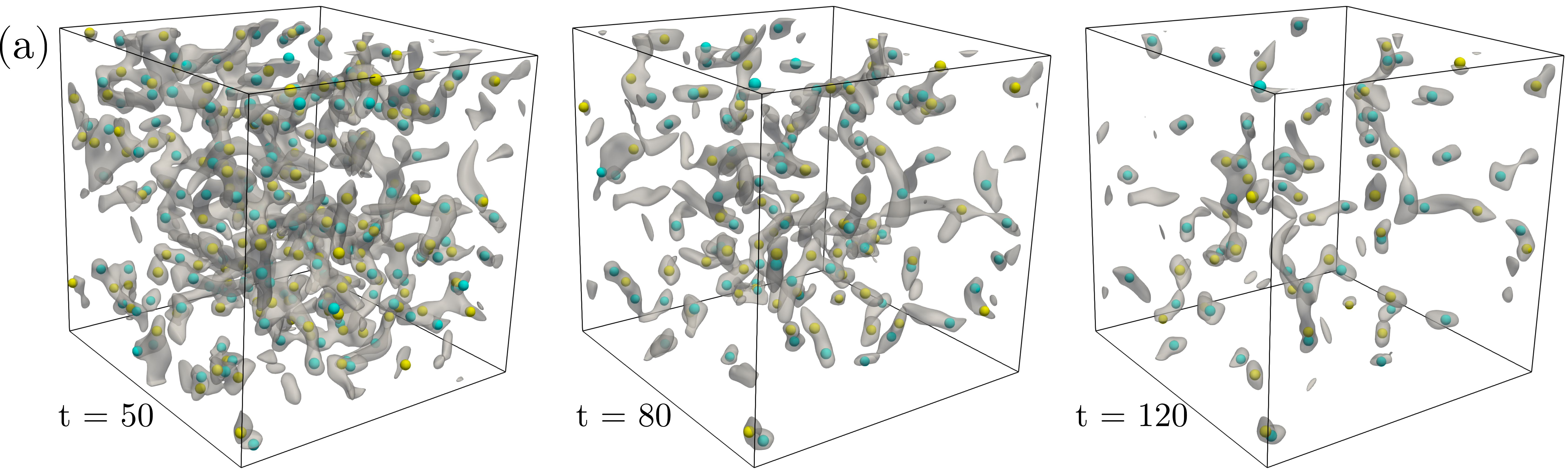}
    \includegraphics[width=\linewidth]{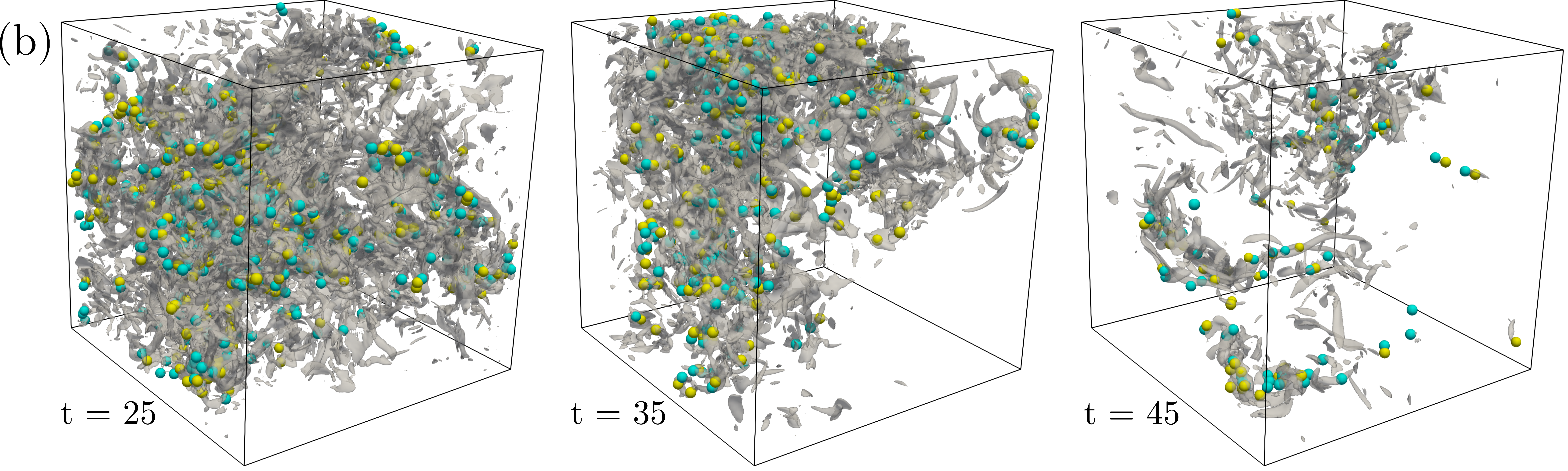}
    \includegraphics[width=\linewidth]{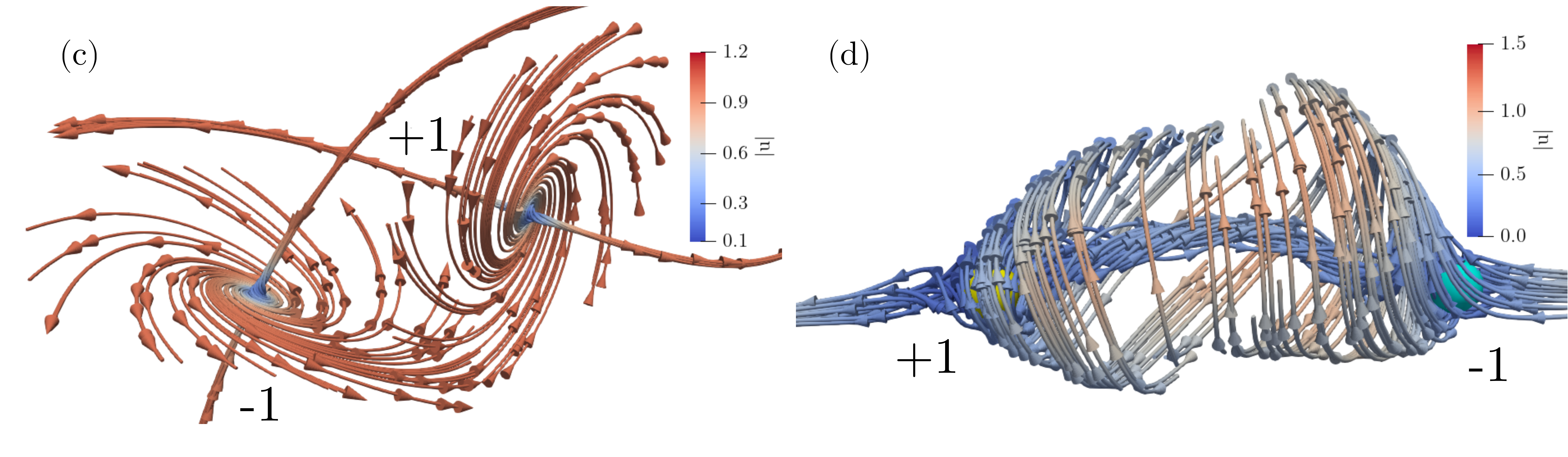}
    \caption{\label{fig:localdissipation}
        Iso-$h$ surfaces overlayed on the defect positions marked by colored spheres (blue : -1, yellow : +1) for (a)
        $\lowre$ and (b) $\highre$ at different stages in the coarsening regime. We show only a subdomain of size
        $(\pi/2)^{3}$ from the simulation box for better visual representation. Streamlines of two neighbour defects
        undergoing merger at (c) $\lowre$ and (d) $\Rey = \pi\times10^{3}$.
    }
\end{figure*}

For the ITT equations, the excess free-energy per unit volume is $h \equiv \Cn^2 |\nabla \vel|^2/2$ and the defects are
identified as velocity nulls, i.e. spatial locations where $\vel={\bm 0}$. Since $n=D=3$ for us, the ITT equation
permits unit magnitude topological charge. We locate defects using the algorithm prescribed by \citet{berg1981} that
has been successfully used to study: (a) the role of defects in the 3D Heisenberg transition \cite{lau1989}, and (b)
coarsening dynamics in the 3D Ginzburg-Landau equations \cite{toyoki1990}. Similar algorithms have also been used to
identify vector nulls in magnetohydrodynamics~\cite{greene1990}, and fluid turbulence \cite{mora2021}.
In~\cref{fig:localdissipation}(a,b) we show the time evolution of the iso-$h$ surfaces overlaid with defect positions
during phase-ordering for low and high-$\Rey$. The streamline plots of pair of oppositely charged defects undergoing
merger are shown in \cref{fig:localdissipation}(c) [$\Rey=0$] and \cref{fig:localdissipation}(d) [$\Rey=2\pi \times
10^4$].

At low $\Rey$, in \cref{fig:localdissipation}(a), we observe that the isosurfaces are primarily localized around the
lines joining oppositely charged defects.  The evolution resembles phase-ordering in the 3D Ginzburg-Landau
equations~\cite{ostlund1981,toyoki1990}. 

In contrast, at high-$\Rey$ we find tubular structures similar to fluid turbulence~\cite{kaneda2003} and the defects
reside in the proximity of these tubes [see \cref{fig:localdissipation}(b)]. Furthermore, the visible clustering of
defects at high-$\Rey$ is consistent with the observed clustering of vector nulls in fluid turbulence~\cite{mora2021}.
 
\begin{figure}[!h]
    \centering
    \includegraphics[width=\linewidth]{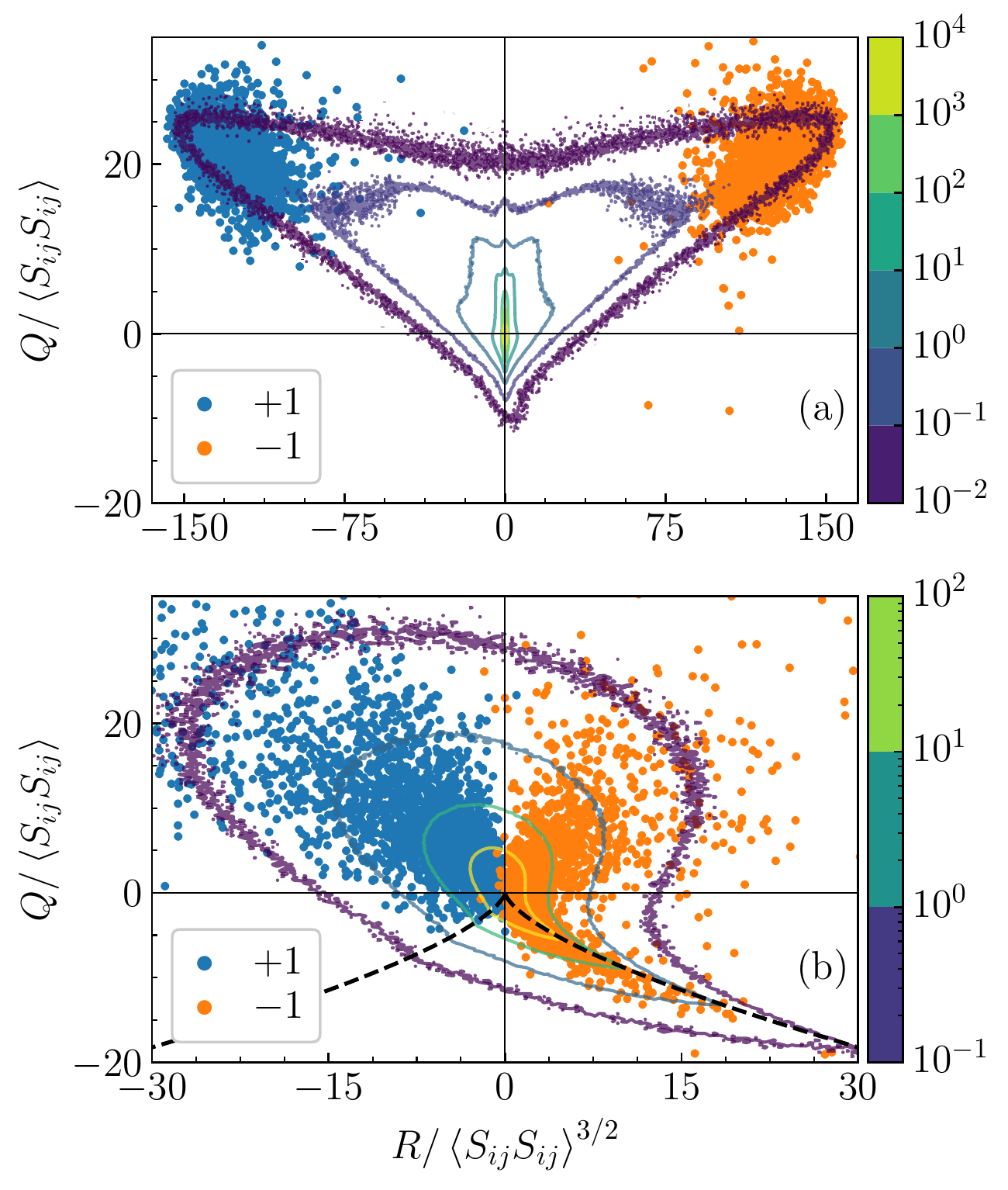}
    \caption{ \label{fig:snap}
        Contour plot of the joint probability distribution $P(R,Q)$ for (a) $\lowre$, and (b) $\highre$ in the
        coarsening regime. Red (Blue) circles mark the position of +1(-1) defect on the $R-Q$ plane. In (b), the black
        dashed line shows the zero-discriminant curve  $D\equiv27R^{2} + 4Q^{3}=0$ that distinguishes vortical ($D>0$)
        and strain-dominated ($D<0$) regions.
    }
\end{figure}

The spatial structures of fluid flows are often characterized by the invariants $Q\equiv -\Tr(\bm{A}^2)/3$ and $R\equiv
-\Tr(\bm{A}^3)/3$ of the velocity gradient tensor $\bm{A} \equiv \nabla\vel$. For high-$\Rey$ fluid turbulence, the
joint probability distribution function $P(R,Q)$ resembles an inverted tear-drop \cite{pandit2009}. In the R-Q plane,
regions above the curve $(27/4) R^2 + Q^3=0$ are vortical, whereas those below are extensional \cite{ooi1999}. From the
flow structures around topological defects [see \cref{fig:localdissipation}(c,d)], it is easy to identify that a
positive (negative) topological charge would have $R<0(>0)$.

In \cref{fig:snap}, we plot the joint probability distribution function $P(R,Q)$ for $\Rey=0$ and $\Rey=2\pi \times
10^4$ at a representative time in the phase-ordering regime. By overlaying the $Q$ and $R$ values at the location of
topological defects on the $P(R,Q)$, as expected, we find that the negative (positive) defects occupy the region with
$R>0(<0)$. For $\Rey=0$, we find symmetric $P(R,Q)$ located primarily in the region $Q>0$, indicating that the flow
structures are vortical. In contrast, for $\Rey=2\pi \times 10^4$ we observe that $P(R,Q)$ has a tear-drop shape
reminiscent of fluid turbulence. The tail region ($Q>0$ and $R<0$) indicates strongly dissipative extensional flow regions
(which also carry a negative charge).  


A unique length scale typically describes the dynamics of systems undergoing phase-ordering, this is often referred to
as the dynamic scaling hypothesis. In the following sections we investigate the validity of this hypothesis for
phase-ordering in the ITT equation for low and high-$\Rey$.

\begin{figure}[!h]
    \centering
    \includegraphics[width=0.8\linewidth]{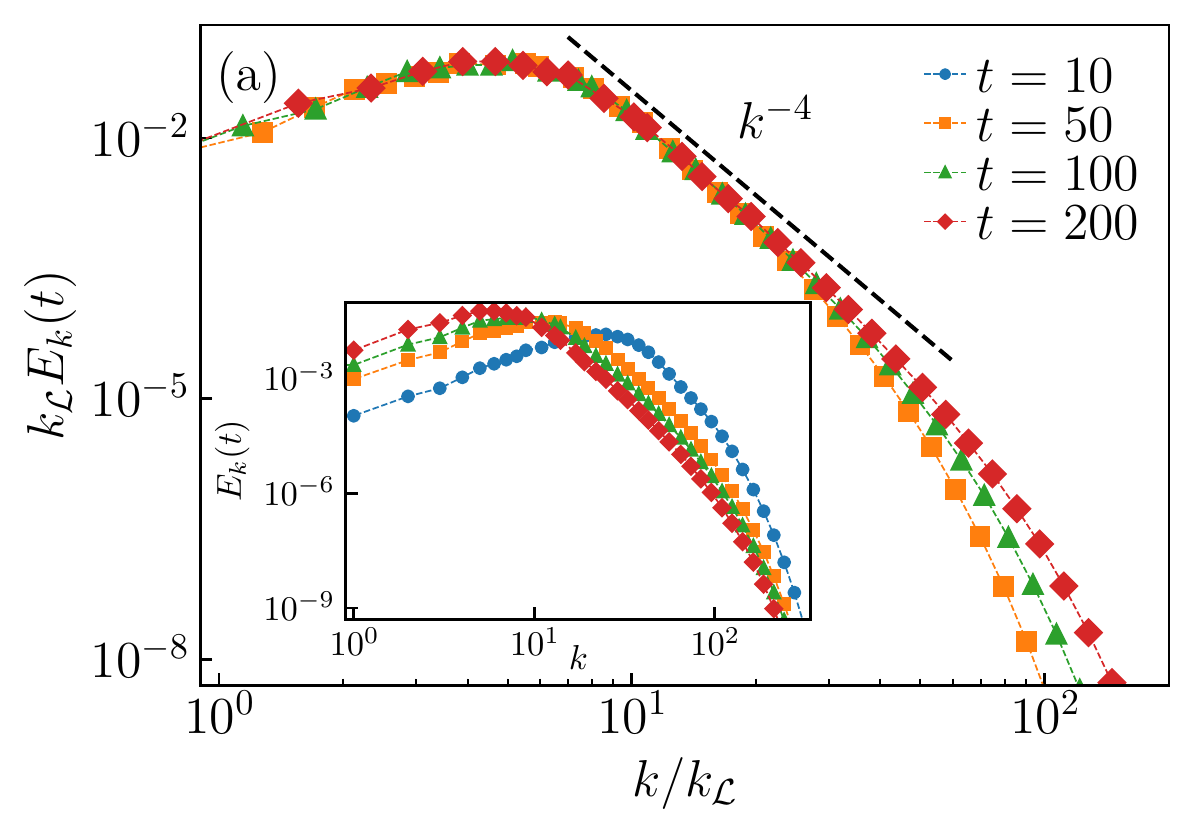}
    \includegraphics[width=0.8\linewidth]{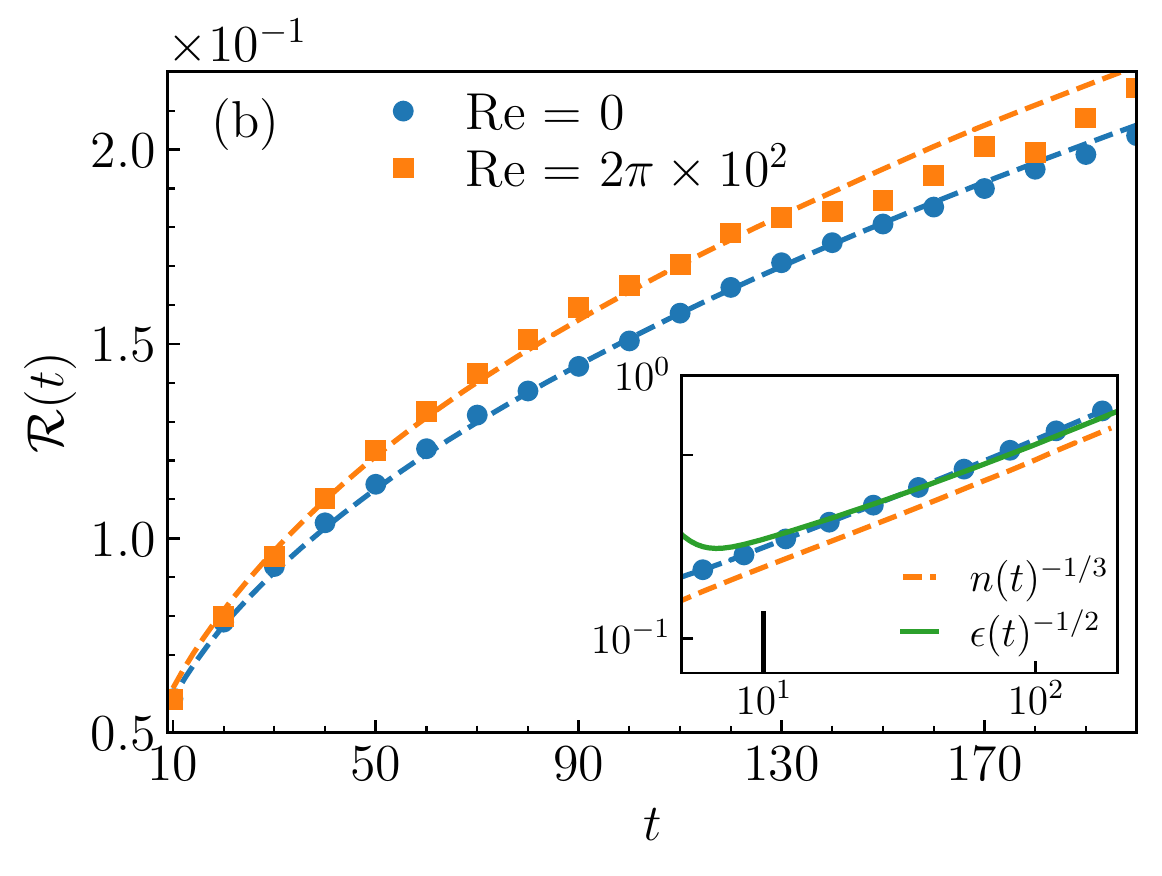}
    \caption{ \label{fig:rt_lowre}
        (a) Scaled energy spectrum $k_{\cal L} E_k(t)$ versus $k/k_{\cal L}$ for $\Rey=0$ at different times. For $k \ll
        k_{\mathcal L}$, we observe Porod's scaling $E_k(k)\sim k^{-4}$. (Inset) Time evolution of the energy spectrum.
        (b) Evolution of average minimum inter-defect separation $R(t)$ for $\Rey=0$ and $\Rey=2\pi \times 10^{2}$.
        Dashed lines show the evolution of  $\mathcal{L}(t)$ (scaled for comparison with $R(t)$). (Inset) Plot showing
        ${\mathcal L}(t) n(t)^{1/3}$ versus $t$ at $\Rey=0$.
    }
\end{figure}

{\it Low Reynolds number--} In \cref{fig:rt_lowre}, we  plot the energy spectrum.  With time, the peak of the spectrum
shifts towards small-wave numbers, indicating a growing length scale often defined as
\cite{onuki2002,qian2003,bray1993,yurke1993,mondello1990,onuki2002,rana2020}
\begin{equation}
    {\mathcal L}(t) \equiv 2 \pi \frac{\sum_k E_k(t)}{\sum_k k E_k(t)}.
\end{equation}
For low $\Rey \to 0$, consistent with the Ginzburg-Landau scaling, we observe ${\mathcal L}(t) \sim \sqrt{t}$ [see
\cref{fig:rt_lowre}(b)] \cite{bray2002}. The rescaled energy spectrum $k_{\mathcal L}E_{k}$ versus $k/k_{\mathcal L}$
collapses onto a single curve for different times [see \cref{fig:rt_lowre}(a)], and we observe Porod's scaling $E_{k
{\mathcal L}} \propto (k{\mathcal L})^{-4}$ for $k {\mathcal L}(t) > 1$ due to the presence of defects
\cite{bray2002}.

\begin{figure*}
    \includegraphics[width=0.32\linewidth]{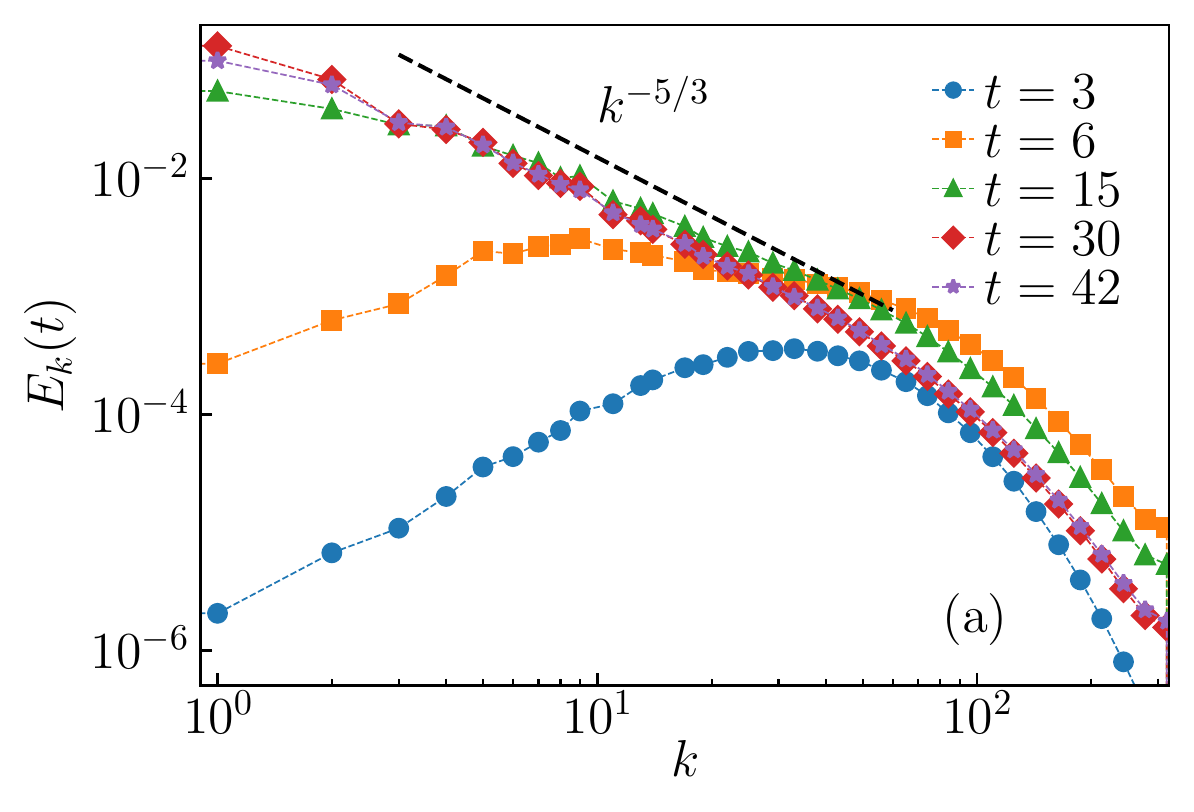}
    \includegraphics[width=0.32\linewidth]{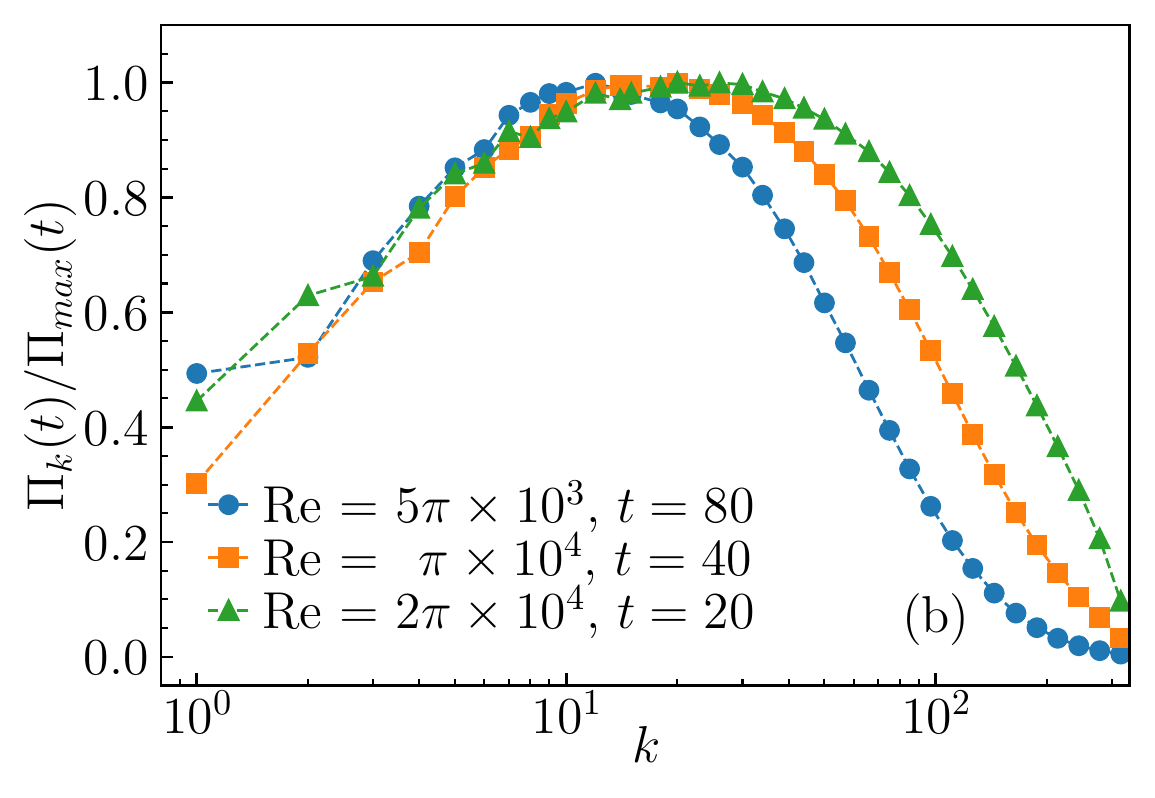}
    \includegraphics[width=0.32\linewidth]{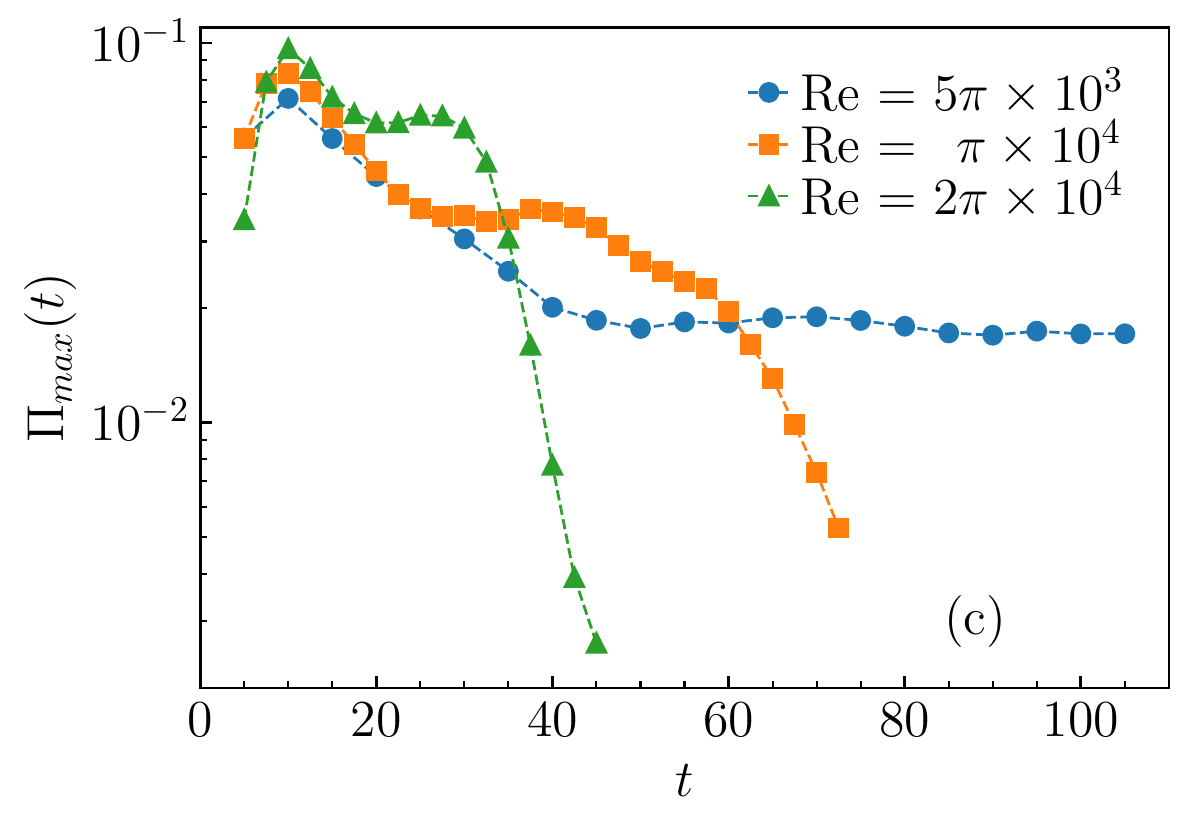}
    \caption{ \label{fig:rt_highre}
        (a) Evolution of the energy spectrum at $\highre$. Dashed black line shows the Kolmogorov scaling $k^{-5/3}$.
        (b) Energy flux at different $\Rey$ in the coarsening regime. At higher $\Rey$, the wavenumber range over which
        we observe the energy flux increases.
        (c) Evolution of the maximum of energy flux $\Pi_{max}(t)$ at various $\Rey$.
    }
\end{figure*}

\begin{figure}
    \centering
    \includegraphics[width=\linewidth]{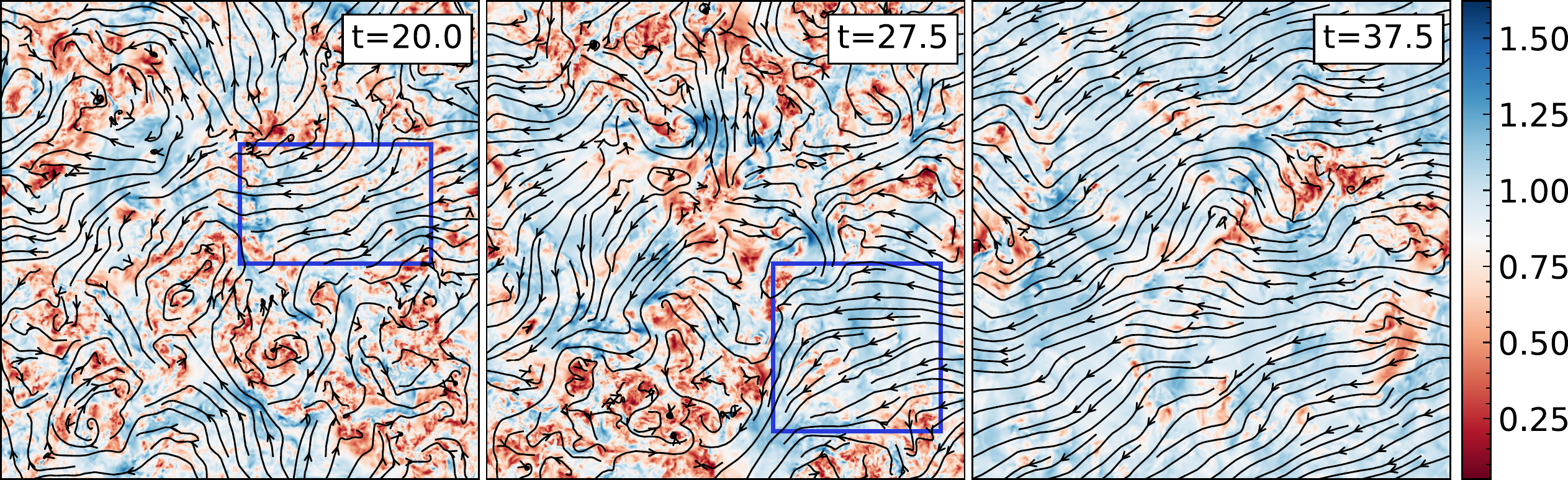}
    \caption{\label{fig:slicehre}
        Pseudocolor plot of the velocity magnitude (z=$\pi$ plane) along with the velocity streamlines during the late
        stages of phase-ordering for $\Rey=2 \pi \times 10^4$.
    }
\end{figure}


The plot in \cref{fig:rt_lowre}(b) shows that the average minimum inter-defect separation $R(t)\propto
{\mathcal L}(t)$, validating the dynamical scaling hypothesis. For uniformly distributed defects, we expect ${\mathcal
L}(t) \propto n(t)^{-1/3}$, where $n(t)$ is the defect number density \cite{chandrasekhar1943,hertz1909,rana2020}. We
verify this in \cref{fig:rt_lowre}(b,inset). 

{\it High Reynolds number--} In \cref{fig:rt_highre}(a) we show the time evolution of the energy spectra for high $\Rey
= 5\pi\times 10^3$. At early times, similar to small $\Rey$, the peak of the spectrum shifts towards small $k$. At
intermediate times which correspond to the plateau region in \cref{fig:localdissipation} (b), we observe a Kolmogorov
scaling $E_k \sim k^{-5/3}$ [\cref{fig:rt_highre}(b)]. A crucial feature of homogeneous, turbulence  is the existence of
a region of constant energy flux $\Pi_k\equiv \lambda \sum_{|\bm{p}| \leq |\bm{k}|}^{N/2}
\widehat{\vel}_{\bm{p}}\cdot(\widehat{\vel\cdot\nabla\vel})_{\bm{-p}}$ \cite{frisch1995}. We evaluate $\Pi_k$ for
different $\Rey=2\pi \times 10^{4}$ at representative times  in the phase-ordering regime and find that it remains
nearly constant between wave-numbers corresponding to the coarsening scale $k_{\mathcal L}\sim 2\pi/{\mathcal L}$ and
the dissipation scale $k_{\eta}\sim(\nu^3/\Pi_k)^{1/4}$ [\cref{fig:rt_highre}(b)].    In \cref{fig:rt_highre}(c),  we
show the time  evolution of $\Pi_{max}(t)\equiv max[\Pi_k(t)]$. The time range over which $\Pi_{max}(t)$ is nearly
constant coincides with the plateau region in ${\mathcal E}(t)$. At high-$\Rey$, the strong turbulence--marked by a
broader inertial range and higher magnitude of $\Pi_{max}$-- leads to a faster ordering. On reducing the $\Rey$, the
cascade range and its strength reduces, we observe a broader time-window over which $\Pi_{max}$ remains constant.

Thus the following picture of phase-ordering emerges: active driving $\alpha - \beta |\vel|^{2}$ injects energy
primarily at large length scales, which is then redistributed to small scales by a forward energy cascade due to the
advective nonlinearity. At late times, we observe regions of turbulence interspersed with growing patches of order
(\cref{fig:slicehre}).

To conclude, we study the coarsening dynamics of the 3D ITT equation. We find that similar to 2D, phase ordering
proceeds via repeated defect merger \cite{rana2020}. At low $\Rey$, the defects are uniformly distributed throughout the
domain and the dynamics is characterized by a unique growing length scale. On the other hand, at high $\Rey$, defects
are clustered and the advective nonlinearities controls the phase ordering. In particular,  we observe the Kolmogorov
scaling in the energy spectrum and a region of constant energy flux.

\begin{acknowledgments}
    We thank G. Garg, and P. B. Tiwari for porting the spectral code onto GPUs. We acknowledge support of the Department
    of Atomic Energy, Government of India, under Project Identification No. RTI 4007.
\end{acknowledgments}


\begin{thebibliography}{35}%
\makeatletter
\providecommand \@ifxundefined [1]{%
 \@ifx{#1\undefined}
}%
\providecommand \@ifnum [1]{%
 \ifnum #1\expandafter \@firstoftwo
 \else \expandafter \@secondoftwo
 \fi
}%
\providecommand \@ifx [1]{%
 \ifx #1\expandafter \@firstoftwo
 \else \expandafter \@secondoftwo
 \fi
}%
\providecommand \natexlab [1]{#1}%
\providecommand \enquote  [1]{``#1''}%
\providecommand \bibnamefont  [1]{#1}%
\providecommand \bibfnamefont [1]{#1}%
\providecommand \citenamefont [1]{#1}%
\providecommand \href@noop [0]{\@secondoftwo}%
\providecommand \href [0]{\begingroup \@sanitize@url \@href}%
\providecommand \@href[1]{\@@startlink{#1}\@@href}%
\providecommand \@@href[1]{\endgroup#1\@@endlink}%
\providecommand \@sanitize@url [0]{\catcode `\\12\catcode `\$12\catcode
  `\&12\catcode `\#12\catcode `\^12\catcode `\_12\catcode `\%12\relax}%
\providecommand \@@startlink[1]{}%
\providecommand \@@endlink[0]{}%
\providecommand \url  [0]{\begingroup\@sanitize@url \@url }%
\providecommand \@url [1]{\endgroup\@href {#1}{\urlprefix }}%
\providecommand \urlprefix  [0]{URL }%
\providecommand \Eprint [0]{\href }%
\providecommand \doibase [0]{http://dx.doi.org/}%
\providecommand \selectlanguage [0]{\@gobble}%
\providecommand \bibinfo  [0]{\@secondoftwo}%
\providecommand \bibfield  [0]{\@secondoftwo}%
\providecommand \translation [1]{[#1]}%
\providecommand \BibitemOpen [0]{}%
\providecommand \bibitemStop [0]{}%
\providecommand \bibitemNoStop [0]{.\EOS\space}%
\providecommand \EOS [0]{\spacefactor3000\relax}%
\providecommand \BibitemShut  [1]{\csname bibitem#1\endcsname}%
\let\auto@bib@innerbib\@empty
\bibitem [{\citenamefont {Bray}(1993)}]{bray1993}%
  \BibitemOpen
  \bibfield  {author} {\bibinfo {author} {\bibfnamefont {A.~J.}\ \bibnamefont
  {Bray}},\ }\href {\doibase 10.1103/PhysRevE.47.228} {\bibfield  {journal}
  {\bibinfo  {journal} {Phys. Rev. E}\ }\textbf {\bibinfo {volume} {47}},\
  \bibinfo {pages} {228} (\bibinfo {year} {1993})}\BibitemShut {NoStop}%
\bibitem [{\citenamefont {Chaikin}\ and\ \citenamefont
  {Lubensky}(1995)}]{chaikin1995}%
  \BibitemOpen
  \bibfield  {author} {\bibinfo {author} {\bibfnamefont {P.~M.}\ \bibnamefont
  {Chaikin}}\ and\ \bibinfo {author} {\bibfnamefont {T.~C.}\ \bibnamefont
  {Lubensky}},\ }\href@noop {} {\emph {\bibinfo {title} {Principles of
  Condensed Matter Physics}}}\ (\bibinfo  {publisher} {{Cambridge University
  Press}},\ \bibinfo {address} {{Cambridge ; New York, NY, USA}},\ \bibinfo
  {year} {1995})\BibitemShut {NoStop}%
\bibitem [{\citenamefont {Ramaswamy}(2010)}]{ramaswamy2010}%
  \BibitemOpen
  \bibfield  {author} {\bibinfo {author} {\bibfnamefont {S.}~\bibnamefont
  {Ramaswamy}},\ }\href {\doibase 10.1146/annurev-conmatphys-070909-104101}
  {\bibfield  {journal} {\bibinfo  {journal} {Annu. Rev. Condens. Matter
  Phys.}\ }\textbf {\bibinfo {volume} {1}},\ \bibinfo {pages} {323} (\bibinfo
  {year} {2010})}\BibitemShut {NoStop}%
\bibitem [{\citenamefont {Marchetti}\ \emph {et~al.}(2013)\citenamefont
  {Marchetti}, \citenamefont {Joanny}, \citenamefont {Ramaswamy}, \citenamefont
  {Liverpool}, \citenamefont {Prost}, \citenamefont {Rao},\ and\ \citenamefont
  {Simha}}]{marchetti2013}%
  \BibitemOpen
  \bibfield  {author} {\bibinfo {author} {\bibfnamefont {M.~C.}\ \bibnamefont
  {Marchetti}}, \bibinfo {author} {\bibfnamefont {J.~F.}\ \bibnamefont
  {Joanny}}, \bibinfo {author} {\bibfnamefont {S.}~\bibnamefont {Ramaswamy}},
  \bibinfo {author} {\bibfnamefont {T.~B.}\ \bibnamefont {Liverpool}}, \bibinfo
  {author} {\bibfnamefont {J.}~\bibnamefont {Prost}}, \bibinfo {author}
  {\bibfnamefont {M.}~\bibnamefont {Rao}}, \ and\ \bibinfo {author}
  {\bibfnamefont {R.~A.}\ \bibnamefont {Simha}},\ }\href {\doibase
  10.1103/RevModPhys.85.1143} {\bibfield  {journal} {\bibinfo  {journal} {Rev.
  Mod. Phys.}\ }\textbf {\bibinfo {volume} {85}},\ \bibinfo {pages} {1143}
  (\bibinfo {year} {2013})}\BibitemShut {NoStop}%
\bibitem [{\citenamefont {Ramaswamy}(2019)}]{ramaswamy2019}%
  \BibitemOpen
  \bibfield  {author} {\bibinfo {author} {\bibfnamefont {S.}~\bibnamefont
  {Ramaswamy}},\ }\href {\doibase 10.1038/s42254-019-0120-9} {\bibfield
  {journal} {\bibinfo  {journal} {Nat. Rev. Phys.}\ }\textbf {\bibinfo {volume}
  {1}},\ \bibinfo {pages} {640} (\bibinfo {year} {2019})}\BibitemShut {NoStop}%
\bibitem [{\citenamefont {Toner}\ and\ \citenamefont {Tu}(1995)}]{toner1995}%
  \BibitemOpen
  \bibfield  {author} {\bibinfo {author} {\bibfnamefont {J.}~\bibnamefont
  {Toner}}\ and\ \bibinfo {author} {\bibfnamefont {Y.}~\bibnamefont {Tu}},\
  }\href {\doibase 10.1103/PhysRevLett.75.4326} {\bibfield  {journal} {\bibinfo
   {journal} {Phys. Rev. Lett.}\ }\textbf {\bibinfo {volume} {75}},\ \bibinfo
  {pages} {4326} (\bibinfo {year} {1995})}\BibitemShut {NoStop}%
\bibitem [{\citenamefont {Chat{\'e}}(2020)}]{chate2020}%
  \BibitemOpen
  \bibfield  {author} {\bibinfo {author} {\bibfnamefont {H.}~\bibnamefont
  {Chat{\'e}}},\ }\href {\doibase 10/gg43bt} {\bibfield  {journal} {\bibinfo
  {journal} {Annu. Rev. Condens. Matter Phys.}\ }\textbf {\bibinfo {volume}
  {11}},\ \bibinfo {pages} {189} (\bibinfo {year} {2020})}\BibitemShut
  {NoStop}%
\bibitem [{\citenamefont {Chen}\ \emph {et~al.}(2015)\citenamefont {Chen},
  \citenamefont {Toner},\ and\ \citenamefont {Lee}}]{chen2015}%
  \BibitemOpen
  \bibfield  {author} {\bibinfo {author} {\bibfnamefont {L.}~\bibnamefont
  {Chen}}, \bibinfo {author} {\bibfnamefont {J.}~\bibnamefont {Toner}}, \ and\
  \bibinfo {author} {\bibfnamefont {C.~F.}\ \bibnamefont {Lee}},\ }\href
  {\doibase 10.1088/1367-2630/17/4/042002} {\bibfield  {journal} {\bibinfo
  {journal} {New J. Phys.}\ }\textbf {\bibinfo {volume} {17}},\ \bibinfo
  {pages} {042002} (\bibinfo {year} {2015})}\BibitemShut {NoStop}%
\bibitem [{\citenamefont {Chen}\ \emph {et~al.}(2016)\citenamefont {Chen},
  \citenamefont {Lee},\ and\ \citenamefont {Toner}}]{chen2016}%
  \BibitemOpen
  \bibfield  {author} {\bibinfo {author} {\bibfnamefont {L.}~\bibnamefont
  {Chen}}, \bibinfo {author} {\bibfnamefont {C.~F.}\ \bibnamefont {Lee}}, \
  and\ \bibinfo {author} {\bibfnamefont {J.}~\bibnamefont {Toner}},\ }\href
  {\doibase 10.1038/ncomms12215} {\bibfield  {journal} {\bibinfo  {journal}
  {Nat. Commun.}\ }\textbf {\bibinfo {volume} {7}},\ \bibinfo {pages} {12215}
  (\bibinfo {year} {2016})}\BibitemShut {NoStop}%
\bibitem [{\citenamefont {Chen}\ \emph {et~al.}(2018)\citenamefont {Chen},
  \citenamefont {Lee},\ and\ \citenamefont {Toner}}]{chen2018}%
  \BibitemOpen
  \bibfield  {author} {\bibinfo {author} {\bibfnamefont {L.}~\bibnamefont
  {Chen}}, \bibinfo {author} {\bibfnamefont {C.~F.}\ \bibnamefont {Lee}}, \
  and\ \bibinfo {author} {\bibfnamefont {J.}~\bibnamefont {Toner}},\ }\href
  {\doibase 10/gg43c5} {\bibfield  {journal} {\bibinfo  {journal} {New J.
  Phys.}\ }\textbf {\bibinfo {volume} {20}},\ \bibinfo {pages} {113035}
  (\bibinfo {year} {2018})}\BibitemShut {NoStop}%
\bibitem [{\citenamefont {Ostlund}(1981)}]{ostlund1981}%
  \BibitemOpen
  \bibfield  {author} {\bibinfo {author} {\bibfnamefont {S.}~\bibnamefont
  {Ostlund}},\ }\href {\doibase 10/fpcqnj} {\bibfield  {journal} {\bibinfo
  {journal} {Phys. Rev. B}\ }\textbf {\bibinfo {volume} {24}},\ \bibinfo
  {pages} {485} (\bibinfo {year} {1981})}\BibitemShut {NoStop}%
\bibitem [{\citenamefont {Toyoki}(1991)}]{toyoki1990}%
  \BibitemOpen
  \bibfield  {author} {\bibinfo {author} {\bibfnamefont {H.}~\bibnamefont
  {Toyoki}},\ }\href@noop {} {\bibfield  {journal} {\bibinfo  {journal} {J.
  Phys. Soc. Jpn.}\ }\textbf {\bibinfo {volume} {60}},\ \bibinfo {pages} {1153}
  (\bibinfo {year} {1991})}\BibitemShut {NoStop}%
\bibitem [{\citenamefont {Lau}\ and\ \citenamefont {Dasgupta}(1989)}]{lau1989}%
  \BibitemOpen
  \bibfield  {author} {\bibinfo {author} {\bibfnamefont {M.~H.}\ \bibnamefont
  {Lau}}\ and\ \bibinfo {author} {\bibfnamefont {C.}~\bibnamefont {Dasgupta}},\
  }\href {\doibase 10.1103/PhysRevB.39.7212} {\bibfield  {journal} {\bibinfo
  {journal} {Phys. Rev. B}\ }\textbf {\bibinfo {volume} {39}},\ \bibinfo
  {pages} {7212} (\bibinfo {year} {1989})}\BibitemShut {NoStop}%
\bibitem [{\citenamefont {Sinha}\ and\ \citenamefont {Roy}(2010)}]{sinha2010}%
  \BibitemOpen
  \bibfield  {author} {\bibinfo {author} {\bibfnamefont {S.}~\bibnamefont
  {Sinha}}\ and\ \bibinfo {author} {\bibfnamefont {S.~K.}\ \bibnamefont
  {Roy}},\ }\href {\doibase 10.1103/PhysRevE.81.041120} {\bibfield  {journal}
  {\bibinfo  {journal} {Phys. Rev. E}\ }\textbf {\bibinfo {volume} {81}},\
  \bibinfo {pages} {041120} (\bibinfo {year} {2010})}\BibitemShut {NoStop}%
\bibitem [{\citenamefont {Mishra}\ \emph {et~al.}(2010)\citenamefont {Mishra},
  \citenamefont {Baskaran},\ and\ \citenamefont {Marchetti}}]{mishra2010}%
  \BibitemOpen
  \bibfield  {author} {\bibinfo {author} {\bibfnamefont {S.}~\bibnamefont
  {Mishra}}, \bibinfo {author} {\bibfnamefont {A.}~\bibnamefont {Baskaran}}, \
  and\ \bibinfo {author} {\bibfnamefont {M.~C.}\ \bibnamefont {Marchetti}},\
  }\href {\doibase 10.1103/PhysRevE.81.061916} {\bibfield  {journal} {\bibinfo
  {journal} {Phys. Rev. E}\ }\textbf {\bibinfo {volume} {81}},\ \bibinfo
  {pages} {061916} (\bibinfo {year} {2010})}\BibitemShut {NoStop}%
\bibitem [{\citenamefont {Das}\ \emph {et~al.}(2018)\citenamefont {Das},
  \citenamefont {Mishra},\ and\ \citenamefont {Puri}}]{das2018}%
  \BibitemOpen
  \bibfield  {author} {\bibinfo {author} {\bibfnamefont {R.}~\bibnamefont
  {Das}}, \bibinfo {author} {\bibfnamefont {S.}~\bibnamefont {Mishra}}, \ and\
  \bibinfo {author} {\bibfnamefont {S.}~\bibnamefont {Puri}},\ }\href {\doibase
  10.1209/0295-5075/121/37002} {\bibfield  {journal} {\bibinfo  {journal} {EPL
  (Europhysics Letters)}\ }\textbf {\bibinfo {volume} {121}},\ \bibinfo {pages}
  {37002} (\bibinfo {year} {2018})}\BibitemShut {NoStop}%
\bibitem [{\citenamefont {Rana}\ and\ \citenamefont
  {Perlekar}(2020)}]{rana2020}%
  \BibitemOpen
  \bibfield  {author} {\bibinfo {author} {\bibfnamefont {N.}~\bibnamefont
  {Rana}}\ and\ \bibinfo {author} {\bibfnamefont {P.}~\bibnamefont
  {Perlekar}},\ }\href {\doibase 10.1103/PhysRevE.102.032617} {\bibfield
  {journal} {\bibinfo  {journal} {Phys. Rev. E}\ }\textbf {\bibinfo {volume}
  {102}},\ \bibinfo {pages} {032617} (\bibinfo {year} {2020})}\BibitemShut
  {NoStop}%
\bibitem [{\citenamefont {Chardac}\ \emph {et~al.}(2021)\citenamefont
  {Chardac}, \citenamefont {Hoffmann}, \citenamefont {Poupart}, \citenamefont
  {Giomi},\ and\ \citenamefont {Bartolo}}]{chardac2021}%
  \BibitemOpen
  \bibfield  {author} {\bibinfo {author} {\bibfnamefont {A.}~\bibnamefont
  {Chardac}}, \bibinfo {author} {\bibfnamefont {L.~A.}\ \bibnamefont
  {Hoffmann}}, \bibinfo {author} {\bibfnamefont {Y.}~\bibnamefont {Poupart}},
  \bibinfo {author} {\bibfnamefont {L.}~\bibnamefont {Giomi}}, \ and\ \bibinfo
  {author} {\bibfnamefont {D.}~\bibnamefont {Bartolo}},\ }\href
  {http://arxiv.org/abs/2103.03861} {\  (\bibinfo {year} {2021})},\ \Eprint
  {http://arxiv.org/abs/2103.03861} {arXiv:2103.03861} \BibitemShut {NoStop}%
\bibitem [{\citenamefont {Frisch}\ and\ \citenamefont
  {Kolmogorov}(1995)}]{frisch1995}%
  \BibitemOpen
  \bibfield  {author} {\bibinfo {author} {\bibfnamefont {U.}~\bibnamefont
  {Frisch}}\ and\ \bibinfo {author} {\bibfnamefont {A.~N.}\ \bibnamefont
  {Kolmogorov}},\ }\href@noop {} {\emph {\bibinfo {title} {Turbulence: The
  Legacy of {{A}}.{{N}}. {{Kolmogorov}}}}}\ (\bibinfo  {publisher} {{Cambridge
  University Press}},\ \bibinfo {address} {{Cambridge, [Eng.] ; New York}},\
  \bibinfo {year} {1995})\BibitemShut {NoStop}%
\bibitem [{\citenamefont {Rosales}\ and\ \citenamefont
  {Meneveau}(2005)}]{rosales2005}%
  \BibitemOpen
  \bibfield  {author} {\bibinfo {author} {\bibfnamefont {C.}~\bibnamefont
  {Rosales}}\ and\ \bibinfo {author} {\bibfnamefont {C.}~\bibnamefont
  {Meneveau}},\ }\href {\doibase 10.1063/1.2047568} {\bibfield  {journal}
  {\bibinfo  {journal} {Phys. Fluids}\ }\textbf {\bibinfo {volume} {17}},\
  \bibinfo {pages} {095106} (\bibinfo {year} {2005})}\BibitemShut {NoStop}%
\bibitem [{\citenamefont {Onuki}(2002)}]{onuki2002}%
  \BibitemOpen
  \bibfield  {author} {\bibinfo {author} {\bibfnamefont {A.}~\bibnamefont
  {Onuki}},\ }\href {https://doi.org/10.1017/CBO9780511534874} {\emph {\bibinfo
  {title} {Phase Transition Dynamics}}}\ (\bibinfo  {publisher} {{Cambridge
  University Press}},\ \bibinfo {address} {{Cambridge; New York}},\ \bibinfo
  {year} {2002})\BibitemShut {NoStop}%
\bibitem [{\citenamefont {Cox}\ and\ \citenamefont {Matthews}(2002)}]{cox2002}%
  \BibitemOpen
  \bibfield  {author} {\bibinfo {author} {\bibfnamefont {S.}~\bibnamefont
  {Cox}}\ and\ \bibinfo {author} {\bibfnamefont {P.}~\bibnamefont {Matthews}},\
  }\href {\doibase 10.1006/jcph.2002.6995} {\bibfield  {journal} {\bibinfo
  {journal} {J. Comp. Phys.}\ }\textbf {\bibinfo {volume} {176}},\ \bibinfo
  {pages} {430} (\bibinfo {year} {2002})}\BibitemShut {NoStop}%
\bibitem [{\citenamefont {Bratanov}\ \emph {et~al.}(2015)\citenamefont
  {Bratanov}, \citenamefont {Jenko},\ and\ \citenamefont
  {Frey}}]{bratanov2015}%
  \BibitemOpen
  \bibfield  {author} {\bibinfo {author} {\bibfnamefont {V.}~\bibnamefont
  {Bratanov}}, \bibinfo {author} {\bibfnamefont {F.}~\bibnamefont {Jenko}}, \
  and\ \bibinfo {author} {\bibfnamefont {E.}~\bibnamefont {Frey}},\ }\href
  {\doibase 10.1073/pnas.1509304112} {\bibfield  {journal} {\bibinfo  {journal}
  {Proc. Natl. Acad. of Sci.}\ }\textbf {\bibinfo {volume} {112}},\ \bibinfo
  {pages} {15048} (\bibinfo {year} {2015})}\BibitemShut {NoStop}%
\bibitem [{\citenamefont {Berg}\ and\ \citenamefont
  {L{\"u}scher}(1981)}]{berg1981}%
  \BibitemOpen
  \bibfield  {author} {\bibinfo {author} {\bibfnamefont {B.}~\bibnamefont
  {Berg}}\ and\ \bibinfo {author} {\bibfnamefont {M.}~\bibnamefont
  {L{\"u}scher}},\ }\href {\doibase 10/c6xn47} {\bibfield  {journal} {\bibinfo
  {journal} {Nucl. Phys. B}\ }\textbf {\bibinfo {volume} {190}},\ \bibinfo
  {pages} {412} (\bibinfo {year} {1981})}\BibitemShut {NoStop}%
\bibitem [{\citenamefont {Greene}(1990)}]{greene1990}%
  \BibitemOpen
  \bibfield  {author} {\bibinfo {author} {\bibfnamefont {J.~M.}\ \bibnamefont
  {Greene}},\ }in\ \href@noop {} {\emph {\bibinfo {booktitle} {Topological
  Fluid Dynamics}}},\ \bibinfo {editor} {edited by\ \bibinfo {editor}
  {\bibfnamefont {H.}~\bibnamefont {Moffatt}}\ and\ \bibinfo {editor}
  {\bibfnamefont {A.}~\bibnamefont {Tsinober}}}\ (\bibinfo  {publisher}
  {Cambridge University Press},\ \bibinfo {address} {Cambridge},\ \bibinfo
  {year} {1990})\ p.\ \bibinfo {pages} {478}\BibitemShut {NoStop}%
\bibitem [{\citenamefont {Mora}\ \emph {et~al.}(2021)\citenamefont {Mora},
  \citenamefont {Bourgoin}, \citenamefont {Mininni},\ and\ \citenamefont
  {Obligado}}]{mora2021}%
  \BibitemOpen
  \bibfield  {author} {\bibinfo {author} {\bibfnamefont {D.~O.}\ \bibnamefont
  {Mora}}, \bibinfo {author} {\bibfnamefont {M.}~\bibnamefont {Bourgoin}},
  \bibinfo {author} {\bibfnamefont {P.~D.}\ \bibnamefont {Mininni}}, \ and\
  \bibinfo {author} {\bibfnamefont {M.}~\bibnamefont {Obligado}},\ }\href
  {\doibase 10/gh7m6p} {\bibfield  {journal} {\bibinfo  {journal} {Phys. Rev.
  Fluids}\ }\textbf {\bibinfo {volume} {6}},\ \bibinfo {pages} {024609}
  (\bibinfo {year} {2021})}\BibitemShut {NoStop}%
\bibitem [{\citenamefont {Kaneda}\ \emph {et~al.}(2003)\citenamefont {Kaneda},
  \citenamefont {Ishihara}, \citenamefont {Yokokawa}, \citenamefont {Itakura},\
  and\ \citenamefont {Uno}}]{kaneda2003}%
  \BibitemOpen
  \bibfield  {author} {\bibinfo {author} {\bibfnamefont {Y.}~\bibnamefont
  {Kaneda}}, \bibinfo {author} {\bibfnamefont {T.}~\bibnamefont {Ishihara}},
  \bibinfo {author} {\bibfnamefont {M.}~\bibnamefont {Yokokawa}}, \bibinfo
  {author} {\bibfnamefont {K.}~\bibnamefont {Itakura}}, \ and\ \bibinfo
  {author} {\bibfnamefont {A.}~\bibnamefont {Uno}},\ }\href {\doibase
  10.1063/1.1539855} {\bibfield  {journal} {\bibinfo  {journal} {Phys. Fluids}\
  }\textbf {\bibinfo {volume} {15}},\ \bibinfo {pages} {L21} (\bibinfo {year}
  {2003})}\BibitemShut {NoStop}%
\bibitem [{\citenamefont {Pandit}\ \emph {et~al.}(2009)\citenamefont {Pandit},
  \citenamefont {Perlekar},\ and\ \citenamefont {Ray}}]{pandit2009}%
  \BibitemOpen
  \bibfield  {author} {\bibinfo {author} {\bibfnamefont {R.}~\bibnamefont
  {Pandit}}, \bibinfo {author} {\bibfnamefont {P.}~\bibnamefont {Perlekar}}, \
  and\ \bibinfo {author} {\bibfnamefont {S.~S.}\ \bibnamefont {Ray}},\ }\href
  {\doibase 10.1007/s12043-009-0096-6} {\bibfield  {journal} {\bibinfo
  {journal} {Pramana}\ }\textbf {\bibinfo {volume} {73}},\ \bibinfo {pages}
  {157} (\bibinfo {year} {2009})}\BibitemShut {NoStop}%
\bibitem [{\citenamefont {Ooi}\ \emph {et~al.}(1999)\citenamefont {Ooi},
  \citenamefont {Martin}, \citenamefont {Soria},\ and\ \citenamefont
  {Chong}}]{ooi1999}%
  \BibitemOpen
  \bibfield  {author} {\bibinfo {author} {\bibfnamefont {A.}~\bibnamefont
  {Ooi}}, \bibinfo {author} {\bibfnamefont {J.}~\bibnamefont {Martin}},
  \bibinfo {author} {\bibfnamefont {J.}~\bibnamefont {Soria}}, \ and\ \bibinfo
  {author} {\bibfnamefont {M.~S.}\ \bibnamefont {Chong}},\ }\href {\doibase
  10.1017/S0022112098003681} {\bibfield  {journal} {\bibinfo  {journal} {J.
  Fluid Mech.}\ }\textbf {\bibinfo {volume} {381}},\ \bibinfo {pages} {141}
  (\bibinfo {year} {1999})}\BibitemShut {NoStop}%
\bibitem [{\citenamefont {Qian}\ and\ \citenamefont
  {Mazenko}(2003)}]{qian2003}%
  \BibitemOpen
  \bibfield  {author} {\bibinfo {author} {\bibfnamefont {H.}~\bibnamefont
  {Qian}}\ and\ \bibinfo {author} {\bibfnamefont {G.~F.}\ \bibnamefont
  {Mazenko}},\ }\href {\doibase 10.1103/PhysRevE.68.021109} {\bibfield
  {journal} {\bibinfo  {journal} {Phys. Rev. E}\ }\textbf {\bibinfo {volume}
  {68}},\ \bibinfo {pages} {021109} (\bibinfo {year} {2003})}\BibitemShut
  {NoStop}%
\bibitem [{\citenamefont {Yurke}\ \emph {et~al.}(1993)\citenamefont {Yurke},
  \citenamefont {Pargellis}, \citenamefont {Kovacs},\ and\ \citenamefont
  {Huse}}]{yurke1993}%
  \BibitemOpen
  \bibfield  {author} {\bibinfo {author} {\bibfnamefont {B.}~\bibnamefont
  {Yurke}}, \bibinfo {author} {\bibfnamefont {A.~N.}\ \bibnamefont
  {Pargellis}}, \bibinfo {author} {\bibfnamefont {T.}~\bibnamefont {Kovacs}}, \
  and\ \bibinfo {author} {\bibfnamefont {D.~A.}\ \bibnamefont {Huse}},\ }\href
  {\doibase 10.1103/PhysRevE.47.1525} {\bibfield  {journal} {\bibinfo
  {journal} {Phys. Rev. E}\ }\textbf {\bibinfo {volume} {47}},\ \bibinfo
  {pages} {1525} (\bibinfo {year} {1993})}\BibitemShut {NoStop}%
\bibitem [{\citenamefont {Mondello}\ and\ \citenamefont
  {Goldenfeld}(1990)}]{mondello1990}%
  \BibitemOpen
  \bibfield  {author} {\bibinfo {author} {\bibfnamefont {M.}~\bibnamefont
  {Mondello}}\ and\ \bibinfo {author} {\bibfnamefont {N.}~\bibnamefont
  {Goldenfeld}},\ }\href {\doibase 10.1103/PhysRevA.42.5865} {\bibfield
  {journal} {\bibinfo  {journal} {Phys. Rev. A}\ }\textbf {\bibinfo {volume}
  {42}},\ \bibinfo {pages} {5865} (\bibinfo {year} {1990})}\BibitemShut
  {NoStop}%
\bibitem [{\citenamefont {Bray}(2002)}]{bray2002}%
  \BibitemOpen
  \bibfield  {author} {\bibinfo {author} {\bibfnamefont {A.~J.}\ \bibnamefont
  {Bray}},\ }\href {\doibase 10.1080/00018730110117433} {\bibfield  {journal}
  {\bibinfo  {journal} {Advances in Physics}\ }\textbf {\bibinfo {volume}
  {51}},\ \bibinfo {pages} {481} (\bibinfo {year} {2002})}\BibitemShut
  {NoStop}%
\bibitem [{\citenamefont {Chandrasekhar}(1943)}]{chandrasekhar1943}%
  \BibitemOpen
  \bibfield  {author} {\bibinfo {author} {\bibfnamefont {S.}~\bibnamefont
  {Chandrasekhar}},\ }\href {\doibase 10.1103/RevModPhys.15.1} {\bibfield
  {journal} {\bibinfo  {journal} {Rev. Mod. Phys.}\ }\textbf {\bibinfo {volume}
  {15}},\ \bibinfo {pages} {1} (\bibinfo {year} {1943})}\BibitemShut {NoStop}%
\bibitem [{\citenamefont {Hertz}(1909)}]{hertz1909}%
  \BibitemOpen
  \bibfield  {author} {\bibinfo {author} {\bibfnamefont {P.}~\bibnamefont
  {Hertz}},\ }\href@noop {} {\bibfield  {journal} {\bibinfo  {journal} {Math.
  Ann}\ }\textbf {\bibinfo {volume} {67}},\ \bibinfo {pages} {387} (\bibinfo
  {year} {1909})}\BibitemShut {NoStop}%
\end{thebibliography}

%
\end{document}